\documentclass[reprint,unsortedaddress,amsmath,amssymb,aps,pre]{revtex4-2}

\usepackage[utf8]{inputenc}
\usepackage{graphicx}
\usepackage{xcolor}
\usepackage{mathrsfs}

\usepackage{dcolumn}% Align table columns on decimal point
\usepackage{bm}% bold math
\usepackage{hyperref}% add hypertext capabilities
\usepackage{gensymb}
\begin{document}

\title{Self-organized vortex phases and hydrodynamic interactions in Bos taurus sperm cells}% 

\author{Charles R. Packard}\email{charles.robert.packard@emory.edu}
\affiliation{Department of Physics, Emory University, Atlanta, GA, USA}

\author{Shobitha Unnikrishnan}
\affiliation{Department of Physics, North Carolina A\&T State University, Greensboro, NC, USA}
\author{Shiva Phuyal}
\affiliation{Department of Physics, North Carolina A\&T State University, Greensboro, NC, USA}

\author{Soon Hon Cheong}
\affiliation{Department of Clinical Sciences, Cornell University, Ithaca, NY, USA}

\author{M. Lisa Manning}\email{mmanning@syr.edu}
\affiliation{Department of Physics, Syracuse University, Syracuse, NY, USA}
\affiliation{BioInspired Institute, Syracuse University, Syracuse, NY, USA}

\author{Chih-Kuan Tung}\email{ctung@ncat.edu}
\affiliation{Department of Physics, North Carolina A\&T State University, Greensboro, NC, USA}

\author{Daniel M. Sussman}\email{daniel.m.sussman@emory.edu}
\affiliation{Department of Physics, Emory University, Atlanta, GA, USA}

\begin{abstract}
Flocking behavior is observed in biological systems from the cellular to super-organismal length scales, and the mechanisms and purposes of this behavior are objects of intense interest. In this paper, we study the collective dynamics of bovine sperm cells in a viscoelastic fluid. These cells appear not to spontaneously flock, but transition into a long-lived flocking phase after being exposed to a transient ordering pulse of fluid flow. Surprisingly, this induced flocking phase has many qualitative similarities with the spontaneous polar flocking phases predicted by Toner-Tu theory, such as anisotropic giant number fluctuations and non-trivial transverse density correlations, despite the induced nature of the phase and the clearly important role of momentum conservation between the swimmers and the surrounding fluid in these experiments. We also find self-organized global vortex state of the sperm cells, and map out an experimental phase diagram of states of collective motion as a function of cell density and motility statistics. We compare our experiments with a parameter-matched computational model of persistently turning active particles, and find that the experimental order-disorder phase boundary  as a function of cell density and persistence time can be approximately predicted from measures of single-cell properties. Our results may have implications for the evaluation of sample fertility by studying the collective phase behavior of dense groups of swimming sperm.
\end{abstract}

\maketitle

%\tableofcontents

\section{\label{sec::Introduction}Introduction}

Biological microswimmers are known to organize into a rich constellation of dynamical phases -- e.g. bio-turbulent \cite{2012_Wensink}, vortex lattice  \cite{2012_Nagai,xu2024self}, and globally oscillating phases \cite{2017_Chen} of bacterial suspensions -- all resulting from the intrinsic motility of their constituent organisms.
Sperm are a canonical type of microswimmer, and the swimming sperm of different species have been found to self-organize into a variety of macroscopic phases under different conditions \cite{riedel2005self,creppy2015turbulence}. Even more so than bacteria, sperm motility is crucial to their biological function. For example, individual sperm are known to exhibit two distinct patterns of motility, chiral and hyperactive, in response to biochemical signaling common in the female reproductive tract \cite{zaferani2023biphasic}. However, while fertilization is achieved by a single sperm, evidence has recently emerged suggesting that collective dynamics of sperm may be needed for sperm to get close to the fertilization site \cite{Qu2021cooperation,Phuyal2022}. 

\begin{figure*}[t!]
    \centering
    \includegraphics[width=18cm]{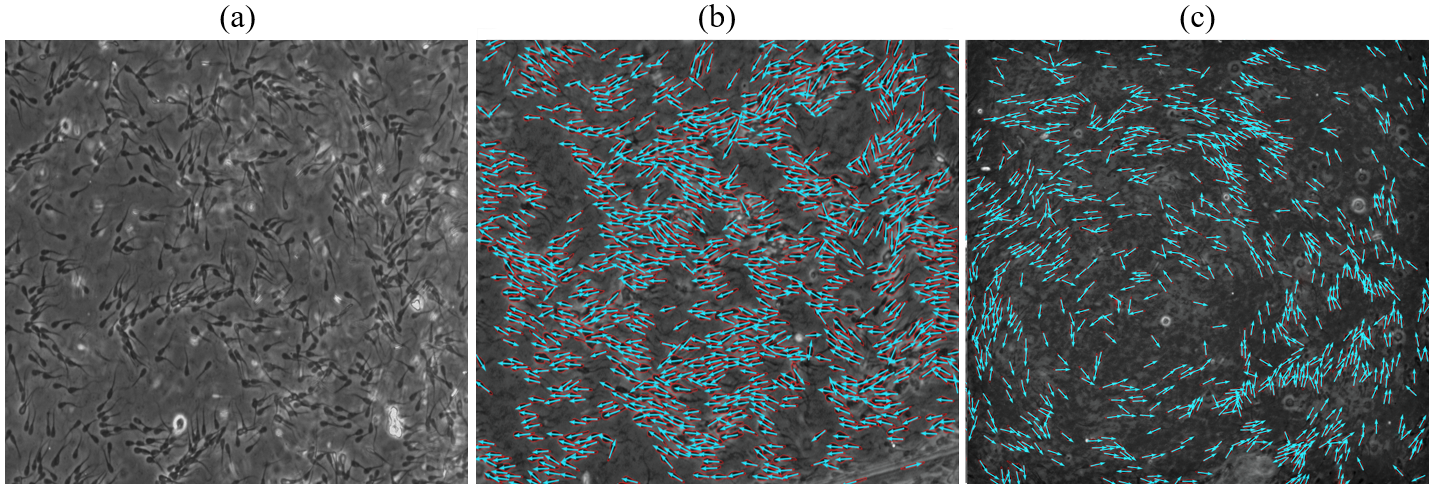}
    \caption{\label{fig::ExpMethods}
    \textbf{Pulse-induced flocking:} Representative images of observed sperm cell phases are shown. (a) Cells swimming in a viscoelastic medium exhibit local polarization, but are globally disordered. After a transient pulse of fluid flow, cells may enter long lived polar flocking (b) or vortex (c) states. In subfigures (b) and (c), red outlines denote automated cell segmentations and blue arrows denote the polar orientations of cells. For scale reference, each cell has a head-to-tail length of $\approx25\mu{m}$ in all images.}
\end{figure*}

In this work we investigate the collective behavior of \textit{Bos taurus} (bovine) sperm cells swimming in a viscoelastic fluid in a microfluidic device at various concentrations subject to a biologically relevant transient hydrodynamic pulse. Surprisingly, we find that even at concentrations for which the sperm do \emph{not} spontaneously form a phase of collective motility, a transient pulse can induce self-organization into at least two different collective phases: polar flocking states and large-scale vortices. These non-equilibrium steady states are difficult to understand both theoretically and via computational models. Modeling these collective dynamics at the level of coarse-grained density and velocity fields -- as in Toner-Tu theories -- is further complicated by the essential role of the solvent in mediating the effective alignment interactions between the cells. It is known that these hydrodynamic effects are important for both the close-range synchronization of sperm flagella and long-range attraction between cells (leading to clustering and lower center-of-mass swimming speed of clusters in some systems) \cite{yang2008cooperation}. In the case of sea urchin spermatozoa, it was shown that detailed hydrodynamic interactions were needed for active matter models to correctly describe an observed vortex lattice phase \cite{yang2014self}. On the other hand, the bio-turbulent phase seen in ram semen \cite{creppy2015turbulence} (which also arises from hydrodynamic interactions \cite{2009_Sokolov, 2015_Lopez}) can also be seen in \textit{B. Subtilis} suspensions and understood by much simpler models which ignore both flagellum and solvent dynamics \cite{2012_Wensink, 2014_GroBmann}.

We are particularly interested in exploring the phenomenology of the long-lived but transiently induced flocking phases we observe, comparing and contrasting these states with models meant to describe standard polar-ordered phases. To what extent do these non-equilibrium states correspond with the single-cell motility statistics of dilute suspensions of cells? Are statistical descriptions of the flocking phases -- for instance, the statistics of density fluctuations and the spatial correlations of those fluctuations -- similar to what is observed standard polar flocking, or are they novel? One might expect that, if the solvent-mediated interactions between sperm cells act only at short length scales within a dense flock,  then the large-scale structure of flocks should be well-described by Toner-Tu hydrodynamics (as, for example, in colloidal rollers \cite{2018_Geyer}). This would imply specific predictions for spatial correlations of density and velocity fluctuations within the flock, such as the presence of long-ranged anisotropic density fluctuation correlations that give rise to giant number fluctuations \cite{toner2019giant} -- we note that the presence of giant number fluctuations has already been observed in some swarming bacteria \cite{zhang2010collective}, although to our knowledge the theoretically predicted anisotropic correlations have not been explicitly studied. It is not a priori obvious to what extent Toner-Tu predictions hold in our system, if at all, given that the theory neglects terms that are relevant here -- e.g., viscoelasticity of the medium, momentum conservation between the swimmers and the medium, etc. -- but our experimental setup creates a novel and useful setting in which to assess the scope of the theory's applicability.

The remainder of this paper is structured as follows. We first describe the experimental conditions under which we observe Bovine sperm cells in a quasi-2-dimensional setting enter either an induced, long-lived vortex or flocking state as a result of a transient hydrodynamic pulse. We then introduce a computational agent-based model to try to understand our experiments, in which collections of self-propelling, aligning, persistently-turning interact in simulation domain with periodic boundary conditions. We use low-density experiments to match parameters between different experiments and our model, and then report the statistical features of the flocking phases observed in our experiments. Surprisingly, we find that the experimental flocking state is characterized by giant number fluctuations, transverse density fluctuation correlations, and anisotropic box scaling statistics that are qualitatively consistent with the predictions of long-wavelength models of active polar liquids \cite{toner2019giant}, although with clear quantitative differences. Finally, we show that the cell density and the persistence time parameter of our agent-based model are key parameters controlling the phase boundary between disordered and ordered phases, allowing us to predict whether transient pulses will lead to collective motion based on measured single-cell properties. This may be of biological relevance regarding the fertilization capacity of different sperm \cite{schoeller2020collective}.

\section{\label{Sec::Methods}Methods}

\subsection{Experimental Procedures and image analyses}
A microfluidic device (2 mm wide and 100 $\mu$m deep channel) was used in conjunction with a syringe pump to provide a well-controlled flow environment, as described in Ref. \cite{Tung2015emergence}. Each device was first filled with a viscoelastic polymer solution, and then equilibrated in an incubator for at least two hours before experiments. During the experiment, the temperature of the microscope stage was kept at 38.5 $\degree$C. Processed sperm suspensions were seeded at the cell seeding opening of the device, and allowed to swim into the device by themselves (see Appendix for diagram). Bovine (\textit{Bos taurus}) sperm have a typical head length and width of $10\mu{m}$ and $5\mu{m}$ respectively, and a tail length of $30\mu{m}$. Due to their natural tendency to swim near a liquid-solid interface \cite{Rothschild1963non, Drescher2011fluid}, most of the sperm swam next to the channel surfaces \cite{Nosrati2015two}, and in this work all images were recorded from the lower surface of the fluidic chamber.
    
Different concentrations of sperm populated the microscope field of view from sample to sample. Even at the highest density, we did not observe a self-organized polar flocking state that formed spontaneously (Fig.~\ref{fig::ExpMethods}a). However, a pulse of viscoelastic fluid flow that transiently aligned the sperm against the flow -- introduced by using a syringe pump set to 1.0 $\mu$L/min for a minute to allow flow rate to ramp up, and then turning off the pump to allow the flow to dissipate over the course of another minute -- was able to induce flocking (Fig.~\ref{fig::ExpMethods}b) with an orientation of the flocks \emph{either} along or against the direction of flow alignment. In addition to flocking states, we observed vortex states in some samples (Fig.~\ref{fig::ExpMethods}c), which were continuously observed for more than 40 minutes.

Quantitative measurements of the ordering of cells in each experimental sample were facilitated by a deep-learning image processing pipeline. We first trained a \textit{cellpose} segmentation model \cite{stringer2021cellpose} on several frames of video data, and then used the model to calculate cell outlines for thousands of additional frames across many different experimental conditions. After training, we compared the positions of sperm cells automatically extracted by this model with a manual annotation of several high-density snapshots (using ImageJ to manually track orientation and location, as well as the Manual Tracking plug-in to compare trajectory information) to verify that the human and deep-learning image annotations were producing the same statistics for the coarse-grained density and orientation fields. Representative segmentations are shown in Fig.~\ref{fig::ExpMethods}b,c.

\subsection{\label{SubSec::ColoredNoiseVicsekModel}Minimal Model For Persistent-Turning Cells}
Density-dependent flocking transitions, like the ones in Fig.~\ref{fig::DensityTunedPhaseTransition}a, 
are characteristic of many models of polar aligning active matter \cite{chate2008collective}, and have been experimentally observed in other cellular systems \cite{szabo2006phase}. However, polar alignment interactions alone though are insufficient to explain the vortex-forming behavior of our low-density samples in Fig.~\ref{fig::DensityTunedPhaseTransition}b. A well-studied alternative, with enough structure to permit vortex phases in principle, are colored-noise ``persistent turning'' models \cite{2012_Nagai,nagai2015collective,sugi2019c,xu2024self}. These models have equations of motion given by
\begin{align}
        \label{Eq::Pos_update}
        \Dot{\mathbf{r}}_i(t)&=v_0\mathbf{e}[\theta_i(t)]
        \\
        \label{Eq::Ori_update}
        \Dot{\theta}_i(t)&=\frac{1}{|\mathcal{N}_i|}
        \sum_{j\in\mathcal{N}_i}\sin(\theta_i-\theta_j) + \omega_i(t)
        \\
        \label{Eq::Omega_update}
        \Dot{\omega}_i(t)&=-\frac{1}{\tau}\omega_i(t)
        + \sqrt{\frac{2}{\tau}}\sigma_{\omega}\zeta_i(t),
\end{align}
where $\mathbf{r}_i$ is the position of particle $i$, and $\mathcal{N}_i$ is the set of neighbors within a radial distance $\ell$ with which the  $i^{th}$ cell aligns its orientation, $\theta_i$. The rotational noise $\omega_i$ obeys an Ornstein-Uhlenbeck (OU) process, with $\zeta_i$ a zero mean, unit variance Gaussian, that is characterized by the persistence time $\tau$ and standard deviation $\sigma_{\omega}$. The correlations $\langle{\omega(t)\omega(t^{\prime})}\rangle\propto e^{-(t-t^{\prime})/\tau}$ of the rotational noise introduces persistent turning to the model, which at sufficiently large $\tau$ and small $\sigma_{\omega}$ may compete with the alignment interactions to produce macroscopic vortex phases \cite{nagai2015collective}.

Equations~\ref{Eq::Pos_update}-\ref{Eq::Omega_update} are a minimal model that we will use to try to capture the complex dynamics of our experimental system. This model approximates the true hydrodynamic interactions between cells, which would favor nematic alignment at long length scales \cite{Tung2015emergence}, as leading to a polar alignment mechanism due to the viscoelastic medium (which dominates the nematic alignment in this system on the length scale of centimeters\cite{tung2017fluid}). Previous studies have found, moreover, that including hydrodynamic interactions is necessary for capturing the collective dynamics at short length and time scales, \cite{riedel2005self,yang2014self}. Here we ask a more basic question, though: given the induced rather than spontaneous nature of our observed flocks, it is not clear that \emph{any} of the standard, highly coarse grained active matter models commonly studied are appropriate for our system. Below we will parameterize this model and explore whether it, nonetheless, has any predictive power in (a) understanding the overall phase behavior we observe in our experiments and (b) helps us understand the detailed statistics of density fluctuations we observe in our flocks.

\section{Characterization of high- and low-density samples}
We first report the overall phase behavior we observe following the transient hydrodynamic pulses described above. In order to classify the dynamics states, we define standard order parameters for the polar and vortex states. For each segmented cell in a given frame, we define the polar orientation of a cell, $\mathbf{n}_{i}$ , from the head-tail asymmetry of the segmentations. We define the global order parameter as $|\langle\mathbf{n}_{i}\rangle|$, where $\langle\cdot\rangle$ refers to the average over all cells in the frame. We defined a vortex order parameter by first calculating the coarse-grained orientation field of the cells, $\mathbf{n}(\mathbf{r})$, and then calculating the vorticity field, $\omega=\nabla\times\mathbf{n}(\mathbf{r})$. We then calculated the enstrophy, $\int{d}\mathbf{r}|\omega|^2$, as a measure of the total amount of vorticity in the frame.

\begin{figure}[ht]
    \centering
    \includegraphics[width=90mm]{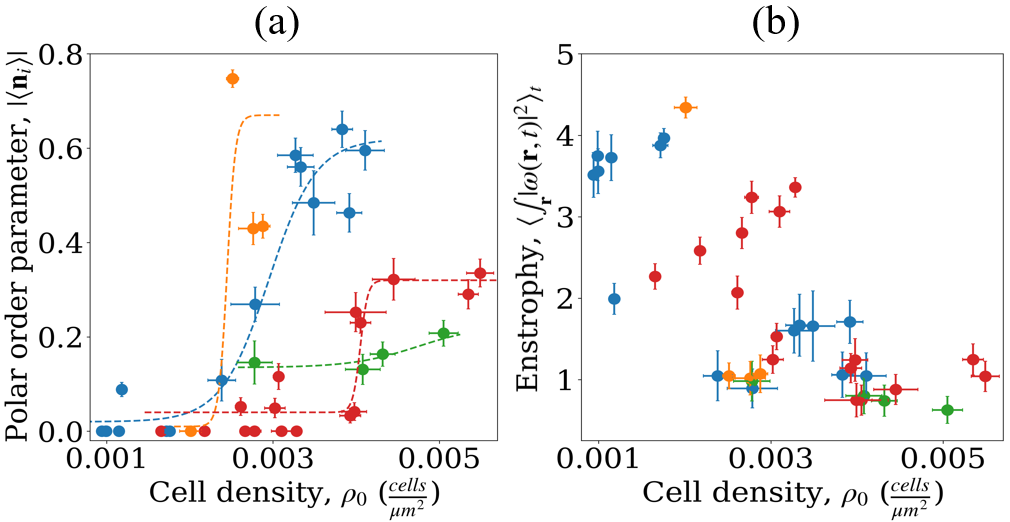}
    \caption{\label{fig::DensityTunedPhaseTransition}
    \textbf{Vortex to flocking transition:} Order parameter values for the (a) polar flocking and (b) and vortex states as a function of cell density are shown for four different experimental samples (denoted by different colors).  Dashed lines in (a) denote fits to hyperbolic tangent functions.}
\end{figure}

Figure \ref{fig::DensityTunedPhaseTransition} show the behavior of these two order parameters as a function of cell density for four different experimental samples taken from different animals. As cell density increases all samples show an increase in the polar order parameter, with three of the four doing so in a way that looks reminiscent of a standard disordered-to-polar-flocking transition (across the range of density we measure, the fourth shows only the beginning of an upturn in the polar order parameter, perhaps indicating that it is in a pre-transition regime). It is clear that, even if this is a transition to a polar-ordered state, the critical density is \emph{not} consistent across samples. We additionally find that the integrated vorticity is anti-correlated with cell density -- i.e., we typically find self-organized vortex states are at generally low cell densities, below the ``critical'' density for the onset of polar order. 
  
\begin{figure}[htb]
    \centering
    \includegraphics[width=90mm]{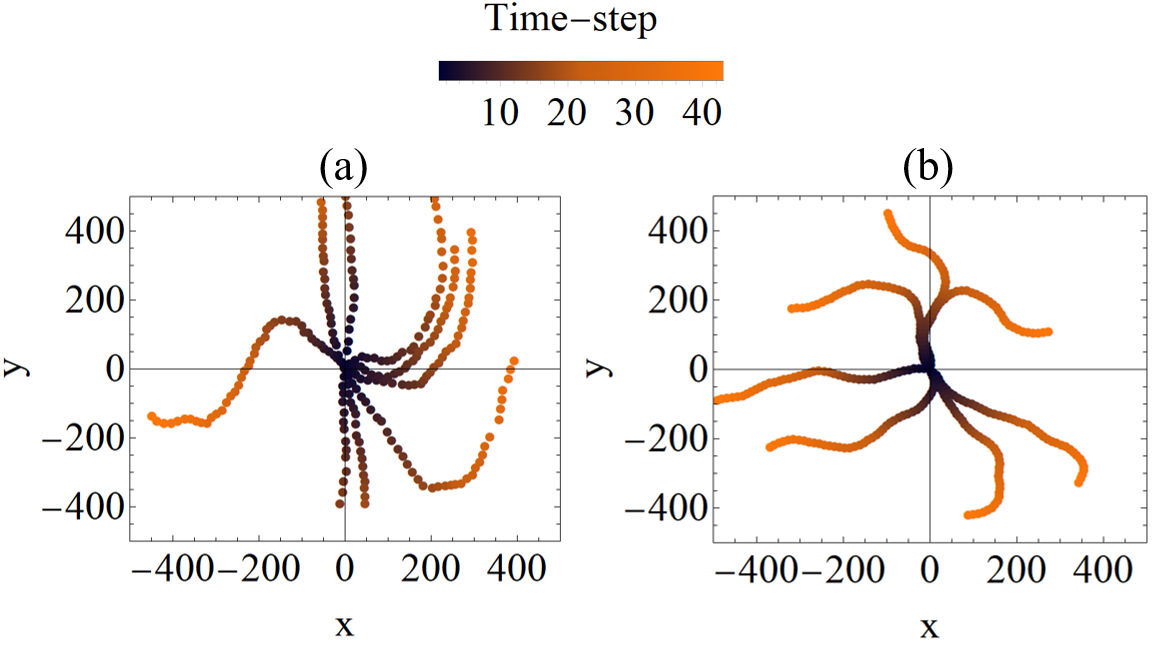}
    \caption{\label{fig::PersistentTurning}
    \textbf{Single cell trajectories:} (a) Representative trajectories of showing single sperm cell displacement from their initial position in a dilute sample. (b) Representative trajectories of non-interacting particles in the persistently-turning Vicsek model for $v_{0}=10$, $\sigma_{\omega}=0.05$, and $\tau=3.0$.}
\end{figure}

\begin{figure}[htb]
    \centering
    \includegraphics[width=90mm]{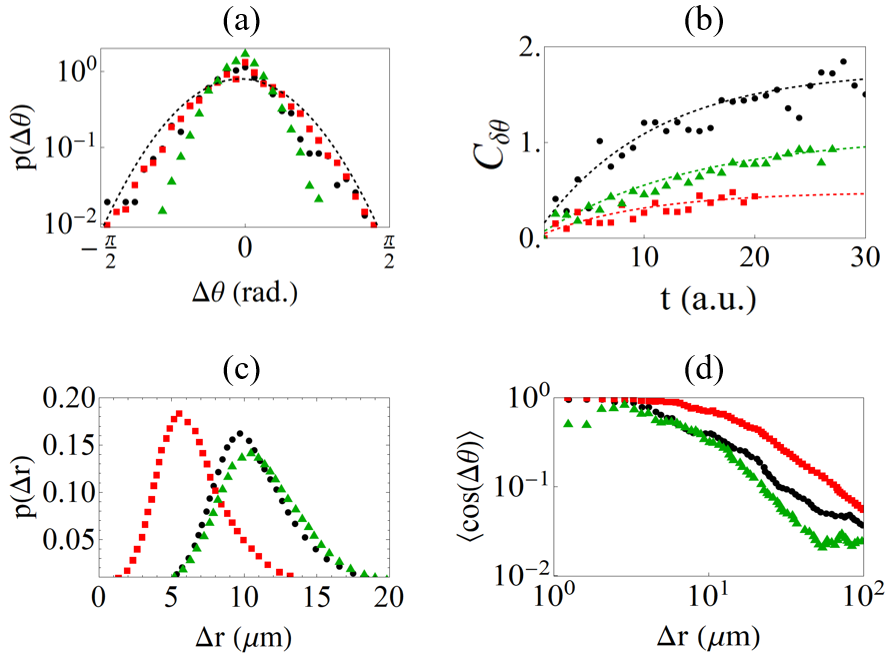}
    \caption{\label{fig::VicsekModelFitting}
    \textbf{Motility statistics for three dilute cell samples used to fit model parameters}. (a) Orientational noise distribution, dashed line denotes normal distribution with zero mean and standard deviation $\sigma=0.50$. (b) Orientation fluctuation correlation function (Eq. \ref{Eq::OrientationFluc_Defn}) computed from $\approx50$ sperm cell trajectories for each sample; dashed lines denote fits of Eq. \ref{Eq::OrientationFluc_CorrFunc}. (c) Spatial displacement distribution between video frames separated by $\sim$150ms. (d) Orientational correlation function. }
\end{figure}

To understand why the critical flocking density may be so different in our different experiments, we investigate the dynamics of individual cells in very dilute concentrations (where cell-cell interactions are minimal). Experimentally, the videos were recorded from the same device filled with the same polymer solution and seeded with the same sperm sample, but at a location further away from the seeding port, therefore lower in cell density, and trajectory tracking mentioned above was used to obtain the statistics. A similar approach was recently employed in Ref. \cite{zaferani2023biphasic}. Figure ~\ref{fig::PersistentTurning} shows a few representative trajectories of sperm at low density, where clear signatures of persistent swimming with a direction that smoothly varies over time are observed. This qualitative observation inspired our use of the persistent-turning dynamics in our computational model described above. From the single-cell statistics shown in Fig.~\ref{fig::VicsekModelFitting} (described in more detail below), it is clear that the microscopic characterization of these different samples are quite different, and in that light it is unsurprising that the onset density varies so substantially. 

In order to more quantitatively compare our experiments with the predictions of the computational model, we measured a variety of single-cell parameters from our experimental samples. A characteristic swimming velocity (corresponding to the self-propulsion speed $v_0$) was estimated by computing the distribution of spatial displacements of cells between frames. As shown in Fig.~\ref{fig::VicsekModelFitting}c, the the swimming speed of isolated cell has a well-defined mean and variance in each sample, and the sample-to-sample difference in mean speeds is within a factor of two. For simplicity, in our later modelling we set the parameter in Eq. \ref{Eq::Pos_update} to the sample mean.

We then measured the distribution of fluctuations in orientation, $\Delta \theta_i$, as shown in Fig.~\ref{fig::VicsekModelFitting}a. We measured the correlation of fluctuations in orientation, $\theta_i(t)$, away from their sample averaged rotation frequency $\omega_0=\langle{\theta_i(T)-\theta_i(0)}\rangle/T$, where $T$ is the total length of time over which cell orientations are measured, via \cite{sugi2019c}
    \begin{equation}
        \label{Eq::OrientationFluc_Defn}
        C_{\delta\theta}(t)=\Big\langle\frac{\theta_i(T)-\theta_i(0)-\omega_0T}{|\theta_i(T)-\theta_i(0)-\omega_0T|}
        \Big(\theta_i(t)-\omega_0t\Big)\Big\rangle\,,
    \end{equation}
If our persistent-turning model of cell motion was correct, the low-density orientational statisticcs should evolve by $\Dot{\theta}_i(t)=\omega_i(t)$ and the orientation fluctuations would be correlated in time as
    \begin{equation}
        \label{Eq::OrientationFluc_CorrFunc}
        C_{\delta\theta}(t)=\sqrt{\frac{2}{\pi}}\sigma_{\omega}\tau\Big(1-e^{-t/\tau}\Big).
    \end{equation}
We measure this correlation function and fit to Eq. \ref{Eq::OrientationFluc_CorrFunc}. As shown in Fig.~\ref{fig::VicsekModelFitting}b we find quite good agreement with this prediction, which we then use to simultaneously estimate $\tau$ and $\sigma_{\omega}$ in our models. Note that Eq.~\ref{Eq::OrientationFluc_CorrFunc} assumes that cells possess Gaussian noise distributions $\omega_{i}$ with a well-defined sample standard deviation $\sigma_{\omega}$; while this is true for some samples, we note that others possess noise distributions with exponential tails (Fig.~\ref{fig::VicsekModelFitting}a). Nevertheless, by treating $\omega_i$ as a Gaussian for all samples, we still find good agreement between Eq. \ref{Eq::OrientationFluc_Defn} and Eq.  \ref{Eq::OrientationFluc_CorrFunc}.

Finally, we investigated whether there were signs of a polar alignment mechanism in these dilute samples by measuring the spatial orientational correlation function of dilute cell trajectories. As shown in Fig.~\ref{fig::VicsekModelFitting}d, in these dilute samples the correlation function exhibits two regimes: a slow initial decay, followed by an intermediate-ranged algebraic decay. These two regimes approximately coincide with the typical size of the sperm cells, with the power-law decay of the latter indicating the presence of hydrodynamically mediated long-ranged interactions. This orientational correlation at low densities is plainly quite different from the kind of finite-ranged aligning interactions assumed in our model, and in the absence of a full calculation of the screening of the hydrodynamic forces in a viscoelastic fluid it is not clear how these orientational statistics would be expected to change at higher densities. Without a characteristic decay length-scale for the orientational correlation function with which to set the interaction cut-off distance for the model, we estimated $\ell$ (the range of polar alignment in our model) as the point at which $\langle{\cos\Delta\theta_{ij}}\rangle$ has decayed by an order of magnitude, and fix $\ell=30\mu{m}$ as the sample mean. This corresponds to just longer than the typical head-to-tail length of the sperm, and while this is reassuringly consistent with the kinds of short-ranged alignment interactions typically assumed, we view these statistics as the weakest agreement with the assumptions of our computational model.

\section{Structure of the Flocking Phase}
In addition to comparing directly with the results of simulating the computational model above, we further compare the character of the induced flocking phase we observe with the qualitative and quantitative predictions made from standard continuum theories of polar active matter. The long-wavelength, long-time properties of polar ordered, self-propelled particles - like the sperm cells in Fig.~\ref{fig::ExpMethods}b - are often described by Toner-Tu theory \cite{1998_Toner}, the full non-linear version of which is still not fully understood \cite{2019_Chate}. While many predictions of the linearized theory have been verified in idealized settings \cite{2018_Geyer}, little is known on how well the predictions of the theory hold when polar alignment interactions are in part mediated by hydrodynamic forces \cite{heidenreich2016hydrodynamic}. In this section we compare the statistics of our flocks with the theory by studying the structure of several correlation functions.

\begin{figure}[h]
        \centering
        \includegraphics[width=90mm]{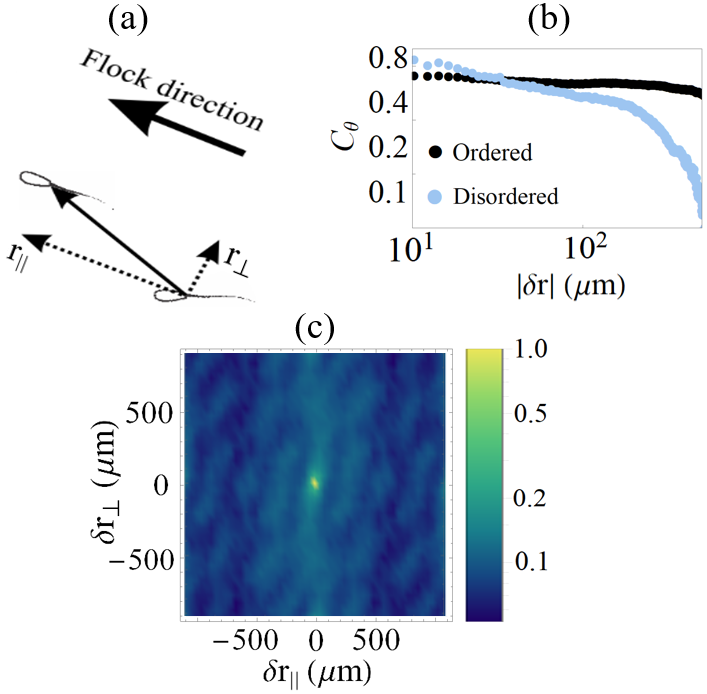}
        \caption{\label{fig::OrientationalSpatialStructure}
        \textbf{Orientational correlations in experiments}
        (a) Coordinate system used for flocking measurements is shown where angles are defined relative to the mean flocking direction, which we use to set the positive $y$-axis. (b) Orientational correlation function of sperm cells in a high-density ordered state (black) and low-density disordered state (blue). (c) Anisotropic orientational correlation function, Eq.~\ref{eq:orientationalCorrelation}, of cells in flocking state.}
\end{figure}

We first study the spatial orientational correlation function,
\begin{equation}\label{eq:orientationalCorrelation}
    C_{\theta}(|\delta{r}|)=\langle{\cos(\theta(r)-\theta(r+|\delta{r}|))}\rangle_{r}\,,
\end{equation}
to establish the long-range order of the flocks we observed after the pulse of flow. Without such a pulse this correlation function quickly decays, but after such a pulse the orientational correlation plateaus at a high value over long distances (Fig.~\ref{fig::OrientationalSpatialStructure}b). This plateau in the correlation function is consistent with our measurement of a large global polar order parameter in identifying the flock. We also note qualitative agreement with a generic prediction of Toner-Tu theory, finding strongly anisotropic orientational correlations: when decomposed along directions parallel and transverse to the flock we find much more rapid decay in the former (Fig.~\ref{fig::OrientationalSpatialStructure}c).

\begin{figure}[h]
        \centering{\includegraphics[width=85.0mm]{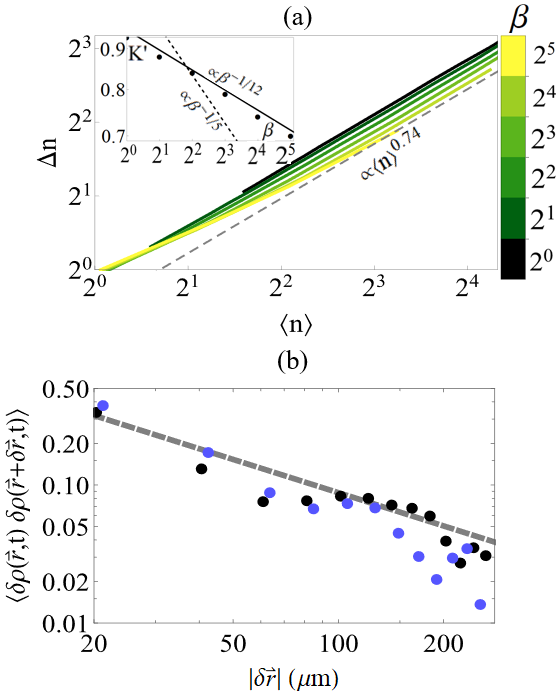}}
        \caption{\label{fig::SpatialStructure}
        \textbf{Density Correlations in experimental flocks:}
        (a) Anisotropic giant number fluctuation scaling for a sperm flock with a polar order parameter of $|\langle\mathbf{n}_{i}\rangle|=0.76$. The inset shows the value of the coefficient $K'$ in Eq.~\ref{eq:tonerScaling} for each value of $\beta$ obtained by power law fits to data in the main plot. (b) Density fluctuation correlation function scaling parallel (black) and transverse (blue) to the global flocking direction. The Toner-Tu scaling predictions $C_{\rho}\propto|\delta{r}|^{-0.8}$ is denoted by the dashed gray line.
       }
\end{figure}

This anisotropy of the flocking state is also revealed in the detailed structure of number fluctuations in the flock. A classic prediction of Toner-Tu theory is that the root-mean-square number fluctuations,  $\Delta{n}\equiv\sqrt{\langle{\delta{n}^2}}\rangle$, should scale as \begin{equation}\label{eq:tonerScaling}
\Delta{n}=K^{\prime}\langle{n}\rangle^{\alpha}.
\end{equation}
We show our measurement of these fluctuations in Fig.~\ref{fig::SpatialStructure}a. The scaling exponent has been predicted to be $\alpha=0.8$ \cite{1998_Toner}, and recent work has suggested that anisotropic correlations in the flock lead to a scaling coefficient that is a function of the aspect ratio of boxes used to count number fluctuations \cite{toner2019giant}. In particular, when the counting boxes are oriented into a coordinate system parallel to and transverse to the mean flocking direction, the prefactor  $K^{\prime}=K^{\prime}(\beta)$ is expected to be a function of the box-shape aspect ratio $\beta=\ell_{||}/\ell_{\perp}$, where in the limit of needle-shaped boxes deep in the flocking phase the prediction is  $K^{\prime}\propto\beta^{-1/5}$. In qualitative agreement with these predictions we observe giant number fluctuations in our induced flocks (albeit with an apparently different exponent). We further observe that the magnitude of the giant number fluctuations strongly depend on the aspect ratio $\beta$, although with a seemingly different exponent (whose low value makes it difficult to distinguish from a potential log correction rather than power-law form). 

We comment, though, that our data is over a relatively limited range of mean densities -- and we are far from the deep-polar-flocking state in which these predictions are expected to hold -- so even ignoring the fact that our phases are induced rather than spontaneous it is not surprising that we observe quantitative discrepancies. Perhaps it is more surprising that these qualitative signatures are nevertheless present in our data. In the appendix we perform a similar analysis on large-scale simulations of Eqs.~\ref{Eq::Pos_update}-\ref{Eq::Omega_update} in a similar parameter regime to the experiment analyzed in Fig.~\ref{fig::SpatialStructure}. We find that a similar power law characterizing the overall density fluctuations over a much larger range of mean densities. This both gives greater confidence in the experimental results shown and reinforces the utility of using a simple model of flocking parameterized according to the low-density motility statistics of individual cells as a potentially predictive tool for our experimental system.

As a final self-consistent measurement of anisotropic flocking statistics, we note that in these anisotropic giant number fluctuations are theoretically expected to result from long-ranged correlations of density fluctuations, $C_{\rho}(\delta\mathbf{r})=\langle{\delta\rho(\mathbf{r})\delta\rho(\mathbf{r}+\delta\mathbf{r})}\rangle$, that decay as $C_{\rho}=|\delta\mathbf{r}|^{-\alpha}G(\theta_{\delta\mathbf{r}})$. Experimentally, we find reasonable agreement as correlations decay with approximately the same exponents both parallel and transverse to the global flocking direction (Fig.~\ref{fig::SpatialStructure}b). However, the correlation functions in the experiments deviate from the predicted exponent of $\alpha=0.8$ at the longest length scales, potentially due to long-ranged hydrodynamic interactions between the cells.  This is consistent the the results in Fig.~\ref{fig::OrientationalSpatialStructure}c, which show that our induced flocking states are inhomogeneous in the longitudinal direction at large length scales. Simulations of our computational model also qualitatively show this long range discrepancy, and suggest that the disagreement may also be due to a strong noise acting on the cells' orientations (i.e., being far from the deeply-ordered polar state, far below the order-disorder transition).

\section{Transitions between Disordered, Polar, and Vortex states}

As noted above, in our experiments we observe sperm collectively forming vortices at cell densities lower than those of the flocks (Fig.~\ref{fig::DensityTunedPhaseTransition}b). We typically see a single vortex composed of hundreds of sperm circling in a field of view larger than 300 $\mu$m, with vortex radii typically larger than $\sim 100\ \mu$m (Fig.~\ref{fig::ExpMethods}c). Although our persistent-turning particle model (Eqs.~\ref{Eq::Pos_update}-\ref{Eq::Omega_update}) is known to possess vortex-forming steady-states, these occur in the \emph{high} density regime of the model \cite{nagai2015collective}. In contrast, the vortex states in our experiments are at fairly low density, and are more similar to those observed in systems of spermatozoa \cite{creppy2016symmetry} and bacteria \cite{wioland2013confinement} confined in a circular geometry, in which chiral symmetry can be broken by hydrodynamic interactions between the cells and the boundary. While our system is not subject to circular confinement (Fig.~\ref{fig::ExperimentalSetup}), it has been known that bovine sperm flagellar beating has uneven handedness \cite{Ishijima1992Rotational}, and the interaction between flagellum and the solid surface leads to the individually chiral trajectories observed here \cite{Friedrich2010high}. Both clearly contribute to the observed vortex yet are not included in the model. From all of this, it is now clear that the simple model will not be able to reproduce the entire phase behavior we observe. Given the results in the previous section -- in which some qualitative features of continuum theories of polar flocks were indeed seen in our data  -- and given that our computational model would coarse grain to a very similar continuum theory, we investigate whether some aspects of the phase transitions seen in the experiments are nevertheless captured.

\begin{figure}[ht!]
        \centering{\includegraphics[width=75.0mm]{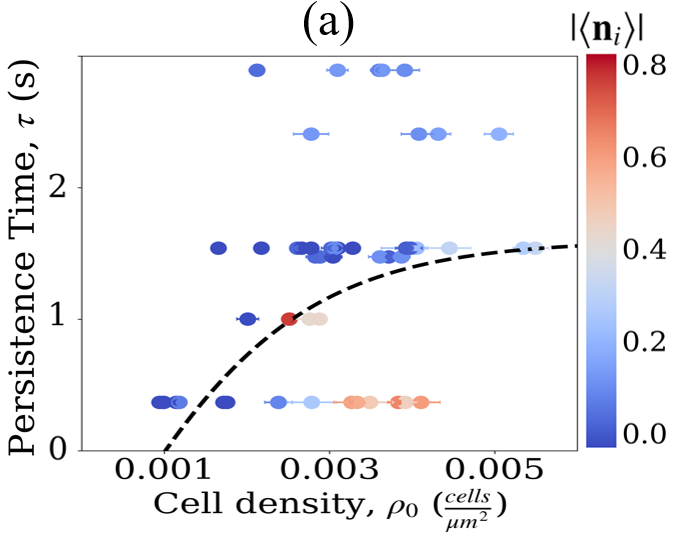}}
        \centering{\includegraphics[width=75.0mm]{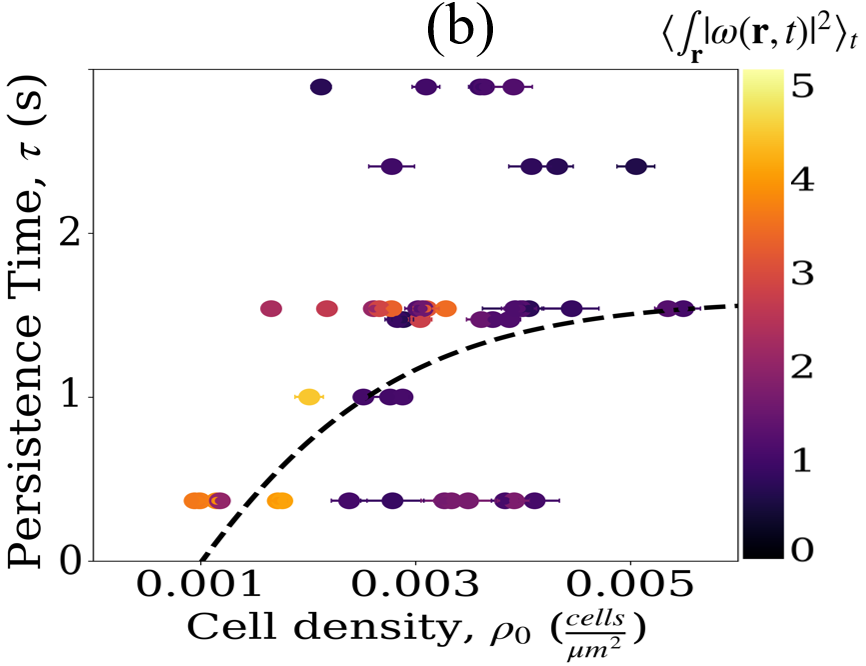}}
        \centering{\includegraphics[width=75.0mm]{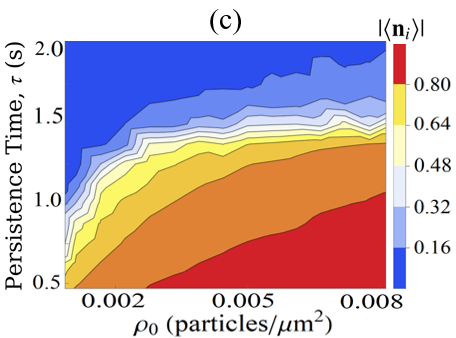}}
        \caption{\label{fig::ExperimentalVsSimulationPhaseDiagrams}
        \textbf{Polar Phase Diagram.}
        Experimental measurement of polar order (a) and enstrophy (b) as a function of cell density and single-cell persistence time. (c) The numerical measurement of the polar order in the persistent turning model at fixed $\sigma_{\omega}=0.25$ for systems of  $N=100,000$ particles. The dashed line in (a) and (b) corresponds to the numerically estimated  phase boundary between disordered and polar ordered steady-states of the persistent-turning model.}
\end{figure}

Cell density is a clear parameter that is predictive of flocking states within individual experimental samples (Fig.~\ref{fig::DensityTunedPhaseTransition}b), and it is theoretically expected that it would play a large role in these transitions. It is clear that other single-cell properties influence the transition and we looked for combinations of model parameters  (corresponding to terms in Eqs. \ref{Eq::Pos_update}-\ref{Eq::Omega_update})  that best predicted whether a system would be in different phases. As shown in Fig.~\ref{fig::ExperimentalVsSimulationPhaseDiagrams}, we find that the combination of global cell density and single-cell rotational persistence time reasonably span the regime over which different macroscopic states are found.

Surprisingly, despite the greatly simplified nature of the cell dynamics in the persistently turning Vicsek model and the limitations we have discussed above, we find that it exhibits an order-disorder transition in a very similar regime of density and persistence time as our experimental system. Specifically, the estimated phase boundary from persistent-tuning simulations (dashed line in Fig 7a,b) appears to be reasonable close to the onset of polar order (7a) and vortex formation (7b) seen in experiments. We certainly anticipate some disagreements between  our model phase boundary and the experiments -- for instance, in our experimental system cells have broad distributions of individual cell properties (c.f. Fig.~\ref{fig::VicsekModelFitting}), and furthermore in the experimental phase diagram many properties (e.g., cell speed $v_{0}$ and rotational noise strength $\sigma_{\omega}$) are varying while in our simulations we vary only $\tau$.

Although the low-density phase of the simple model does not capture the experimentally observed vortex states in the corresponding regions of parameter space, this difference in phase behavior may be attributed to the absence of long-ranged hydrodynamic interactions and confinement effects in the model. In the absence of the screening out of hydrodynamic interactions, polar and weakly nematic alignment at low density may work to stabilize the vortices we observe. Overall, we believe that our results demonstrate that the competition between polar alignment strength, controlled by local density, and persistent turning in the minimal active matter model can predict important aspects of the macroscopic phase behavior of the sperm samples studied here, highlighting the analytical power of the model.

\section{\label{Sec::Discussion}Discussion}

In summary, we report long-lived flocks of bovine sperm induced by transient pulsed flow in a viscoelastic medium. Some aspects of these collective states are reasonably well-described by numerical simulations of a persistently turning variation of the Vicsek model, which takes into account the propensity of isolated sperm cells in this quasi-2D environment to have long-lived orientational correlations in their direction of motion. Surprisingly, we also find giant number fluctuations and transverse density fluctuation correlations that are qualitatively in agreement with the most recent and ``shocking'' predictions of Toner-Tu theories \cite{toner2019giant}.

This is not a trivial comparison and agreement between theory and experiment, as this theoretical description ignores momentum conservation between the self-propelled objects and the surrounding fluid, and the momentum of the fluid is certainly not negligible for microswimmers. It has been shown theoretically that  flocking is, in fact, unstable among low-Reynolds number microswimmers \cite{Simha2002hydrodynamic}. While this conclusion may be modified by the fact that the sperm are swimming very close to the solid substrate \cite{Nosrati2015two} and the no-slip boundary significantly dissipates the movement of fluid, the reduction of flow should also reduce the alignment mechanism at the core of sperm-sperm interactions \cite{Ishimoto2018hydrodynamic}. Another possibility is that there is direct momentum transfer between the sperm and the solid substrate, a hypothesis which has been proposed by biologists for decades \cite{Phillips1972comparative}, but has not been quantitatively verified. Here we provide early evidence strengthening this hypothesis, while more work will be needed to definitively confirm this hypothesis.

Although collective motion is observed in biological systems across many length-scales, the biological relevance or potential function of the different swimming patterns observed here remains unclear. We have studied the collective motion of bovine sperm cells in a viscoelastic medium following a polar ordering pulse -- a combination common in biological fluids -- and find that the polarized cells exhibit long-ranged order for a sustained period of time with density fluctuation statistics consistent with Toner-Tu theory. In natural settings though, sperm cells navigate in three dimensions through far more complex environments to reach the ova. Recent interest in the behavior of active fluids on curved surfaces \cite{sknepnek2015active} -- such as the motion of cells along the gut \cite{ritsma2014intestinal} -- has generated natural extensions of Toner-Tu theory \cite{shankar2017topological} which may provide new insights into the topologically complex environment of the female tract. By analogy with the potential for azimuthal flocking to be regulated by changing Gaussian curvature along the length of an open tube~\cite{shankar2017topological}, we speculate that the vortex phase may have a biological function, perhaps as a reservoir or storage mechanism for the sperm cells.

Our work has direct relevance to the diagnosis of male fertility of some animals in agricultural settings. For human males, the evaluation of sperm motility typically utilizes optical microscopy to visualize the swimming of individual sperm \cite{Gaffney2021modelling}. However, when assessing male fertility for a range of animals in the field, a microscope with environmental control is often not available. In these cases, \emph{mass motility} -- the collective movement of a group of sperm \cite{DAVID2015mass, david2018new} -- has often been used as a quick and easy diagnostic tool for the estimation of male fertility. The highest mass motility scores are associated with macroscopically visible groups of sperm moving in the same direction, similar to the behavior of flocking phases often discussed in the active matter literature. This crude technique has been found to be unreliable though, as sperm concentration and their collective swimming speed are not always positively correlated, and more careful approaches to assessing mass motility are needed \cite{koziol2018manual}. The requirements for the formation of sperm mass motility, or its connection to single-sperm properties, is not well understood. It has been shown to depend both on some intrinsic properties of the sperm cells (such as head shape \cite{breed2005evolution, pearce2018cellular} and flagellum morphology \cite{hook2020methodological}) and on the environment in which the cells are swimming, but many of the details remain unclear. This study therefore helps provide a new method for assessing mass motility by connecting the mass motility scores to the swimming characteristics of individual sperm cells.

\begin{acknowledgments}
This work is supported by NSF HRD 1665004 and NIH R15HD095411 to CKT, American Physical Society Bridge Program Minority Serving Institution Travel Grant to MLM and CKT, and NSF DMR 2143815 to DMS. All bovine semen samples were kindly provided by Genex Cooperative, Inc at Ithaca, NY.
\end{acknowledgments}

\section{Appendix}

\subsection{Media and Semen Sample Preparation}
\begin{figure}[h]
        \centering{\includegraphics[width=90.0mm]{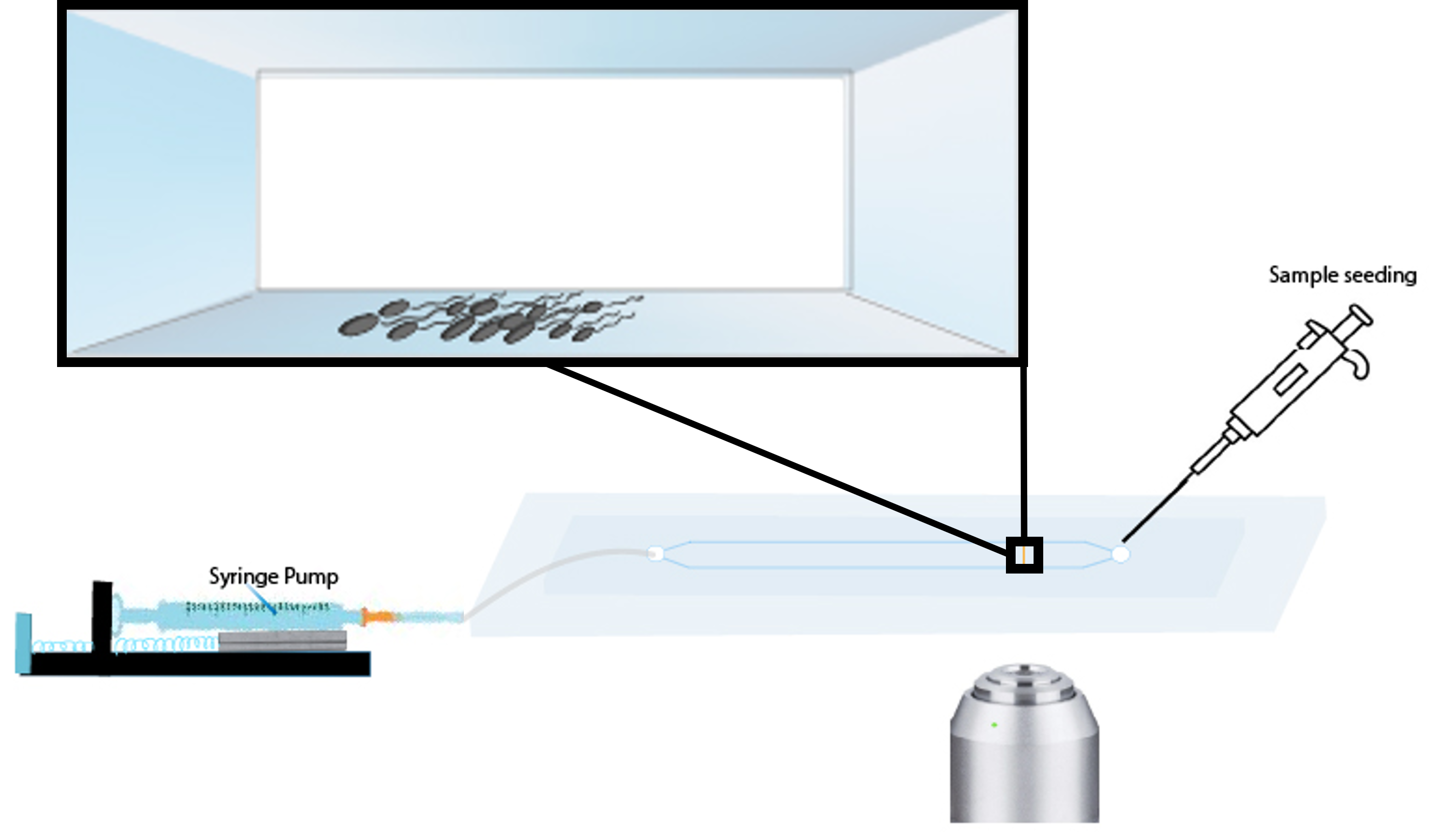}}
        \caption{\label{fig::ExperimentalSetup}
        \textbf{Experimental setup.} A microfluidic device connected to a syringe pump was mounted on an inverted microscope for observation. The syringe pump is used for generation of a controlled flow, while the other end of the channel was left open for seeding sperm into the device. They were initially allowed to swim into the device by themselves, and flocking were seen after a pulse of flow. The upper portion shows an enlarged view of the cross section of the channel, with most of the sperm found on the channel surface.
        }
\end{figure}
Tyrode albumin lactate pyruvate (TALP) \cite{Parrish1988capacitation} was used as sperm medium in the experiments. TALP comprised of 99 mM NaCl, 3.1 mM KCl, 25 mM NaHCO$_3$, 0.39 mM NaH$_2$PO$_4$, 10 mM HEPES free acid, 2 mM CaCl$_2$, 1.1 mM MgCl$_2$, 25.4 mM sodium lactate, 0.11 mg/mL sodium pyruvate, 5 $\mu$g/mL gentamicin, and 6 mg/mL bovine serum albumin (Fraction V; Calbiochem, La Jolla, CA, USA), with a pH of 7.4. To increase the viscoelasticity of the fluid, either 1\% (w/v) of Methyl Cellulose (MC, 4000 cP at 2\%) or a combination of 0.7\% (w/v) of polyacrylamide (PAM, 5-6 MDa) plus 1\% (w/v) of polyvinylpyrrolidone (PVP) are supplemented into TALP. Polymer solutions were made by slowly stirring polymer powders into TALP in room and chilled temperatures alternately. All solutions were equilibrated in a 38.5 Â°C incubator with 5\% CO$_2$ in humidified air until usage. Some samples were fresh ejaculates and some were frozen and stored in liquid nitrogen before being thawed for use. Processing details for both kinds of samples are identical to those described in Ref. \cite{Walker2021computer}. After the processing, sperm suspensions were kept in an incubator maintained at 38.5 Â°C with 5\% of CO$_2$.

\subsection{Density statistics in the agent-based model}
\begin{figure}[ht]
        \centering{\includegraphics[width=85.0mm]{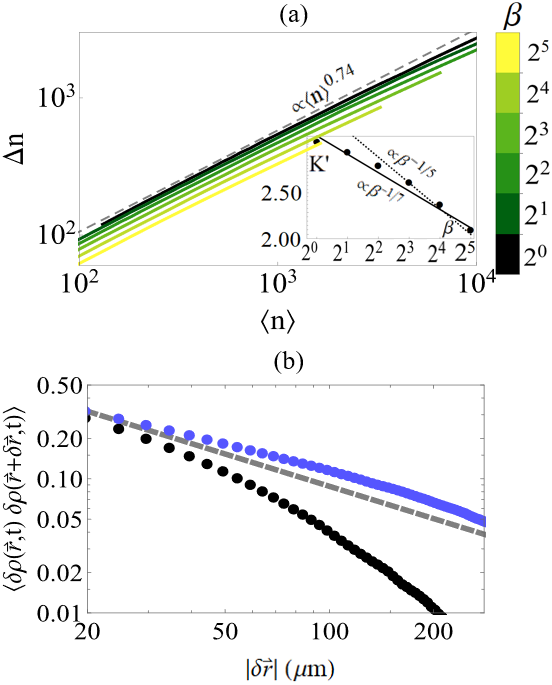}}
        \caption{\label{fig::SpatialStructureComputational}
        \textbf{Density Correlations in computational flocks:}
        (a) Anisotropic giant number fluctuation scaling in simulations of Eqs.~\ref{Eq::Pos_update}-\ref{Eq::Omega_update} with a polar order parameter of $|\langle\mathbf{n}_{i}\rangle|=0.70$. The inset shows the scaling coefficient for each value of $\beta$ obtained by power law fits to data in the main plot. (b) Density fluctuation correlation function scaling parallel (black) and transverse (blue) to the global flocking direction. The Toner-Tu scaling predictions $C_{\rho}\propto|\delta{r}|^{-0.8}$ is denoted by the dashed gray line.
       }
\end{figure}
As a further test of the hypothesis that the simple ``persistent-turning'' flocking model of Eqs.~\ref{Eq::Pos_update}-\ref{Eq::Omega_update} can be used to predict the macroscopic behavior of our experimental system, we analyzed the anisotropic density statistics of $N=10^5$ particles with parameter values $\tau=1.0$, $\sigma_{\omega}=0.25$, $\rho_{0}=2.2\times10^{-3}\,\text{cells}/\mu{m}^{2}$, $v_{0}=7.5\mu{m}$, and $\ell=30\mu{m}$. This leads to a steady state with mean polar order parameter of $|\langle\mathbf{n}_{i}\rangle|=0.70$, which is reasonably close to that observed in the experiments analyzed in Fig.~\ref{fig::SpatialStructure} but still far from the deep-ordered polar phase. 

In Fig.~\ref{fig::SpatialStructureComputational} we report the same density correlation statistics as in the experiments in the main text, but with the ability in the simulations to look over a much broader range of scaling. Surprisingly, we observe overall giant-number fluctuations whose scaling exponent is \emph{quantitatively} consistent with our experimental estimate. Interestingly, the dependence of the scaling prefactor $K'$ on counting box aspect ratio is closer to the theoretical prediction than the experimental result. Finally, the decay of the spatial density correlations parallel and transverse to the flocking directions is notably different from the experimental results: while the transverse correlations seem consistent with both experimental and theoretical results, we observe an unexpectedly rapid decay parallel to the flocking direction. We speculate that the agreement for the transverse correlations may be coincidental, and that for neither of these measurements are in the asymptotically large distance regime for which the theoretical prediction is meant to hold.

\bibliography{flockingBosTaurusPaper}

\end{document}